\documentclass[12pt]{article}
\usepackage[authoryear]{natbib}
\usepackage{graphicx}
\usepackage{amssymb,amsthm,amsmath,epsfig}
\usepackage{times}
\usepackage{bm}
\usepackage{color}
\usepackage{setspace}
\usepackage[utf8]{inputenc}
\usepackage{lscape}
\usepackage[nottoc]{tocbibind}
\usepackage[toc,page]{appendix}

\usepackage[english]{babel}

\usepackage{actuarialangle}
\usepackage{actuarialsymbol}

\newcommand\x{2.5}
\usepackage[top = \x cm, bottom = \x cm, left = \x cm,right = \x cm]{geometry}

\usepackage{changes}
\usepackage{lipsum}
\definechangesauthor[name={Trifon I. Missov}, color=orange]{TIM}

\newcommand{\e}{{\rm e}}

\usepackage{authblk}

\usepackage{fancyhdr}

\title{Using a Penalized Likelihood to Detect Mortality Deceleration}

\author[1]{Silvio C. Patricio\thanks{silca@sam.sdu.dk}}
\author[1]{Trifon I. Missov}
\affil[1]{\small{The Interdisciplinary Centre on Population Dynamics, University of Southern Denmark}}

\begin{document}

\date{}
\maketitle

\begin{abstract}
\noindent{In this paper, we suggest a novel method for detecting mortality deceleration. We focus on the gamma-Gompertz frailty model and suggest the subtraction of a penalty in the log-likelihood function as an alternative to traditional likelihood inference and hypothesis testing. Over existing methods, our method offers advantages, such as avoiding using a p-value, hypothesis testing, and asymptotic distributions. We evaluate the performance of our approach by comparing it with traditional likelihood inference on both simulated and real mortality data. Results have shown that our approach is more accurate in detecting mortality deceleration and provides more reliable estimates of the underlying parameters. The proposed method is a significant contribution to the literature as it offers a powerful tool for analyzing mortality patterns.}\\
\noindent{\em Keywords}: Gompertz model; gamma-Gompertz model, mortality deceleration; penalized likelihood function; maximum a posteriori probability.
\end{abstract}



\section{Introduction}

Human death-rate patterns are astoundingly log-linear over a wide range of adult ages. The Gompertz distribution \citep{gompertz1825xxiv} with an exponentially increasing hazard function captures this accurately. The theory of unobserved heterogeneity and the associated frailty model \citep{vaupel1979impact} predicts a downward deviation at the oldest ages, to which only the most robust individuals in the population survive. Detecting such a deceleration in real data is not always successful \citep{gavrilova2015biodemography, newman2018errors}, even though the vast majority of studies indicate that death rates at older ages increase at lower rates and can even level off \citep{Curetal92, Fuketal93, Fuketal96, Caretal95, Khaetal98, gampe2010human, gampe2021mortality, rootzen2017human, alvarez2021regularities, camarda2022curse, Beletal22}. In a frailty model setting, testing for mortality deceleration is equivalent to testing whether the non-negative frailty parameter is strictly positive.

Formally, denote by $X$ a non-negative continuous random variable that describes individual human lifespans (complete or after a given adult age). If $X \sim$ Gompertz$(a,b)$, where $a$ is the mortality level at the initial age and $b$ is the rate of aging, the associated hazard function (force of mortality) at time $x$

\begin{equation*}
    \mu(x) = \lim_{\varepsilon \downarrow 0} \mathbb{P} (x \leq X < x+\varepsilon | X \geq x).
\end{equation*}
is $\mu(x) = ae^{bx}$ . \cite{vaupel1979impact} introduce a positive continuous random variable $Z$, called \textit{frailty}, that acts multiplicatively on $\mu(x)$ and captures one's unobserved susceptibility to death. The force of mortality for an individual with frailty $Z = z$ is

\begin{equation*}
    \mu(x \, | \, Z=z) = z \, \mu(x) \,.
\end{equation*}
For a gamma-distributed frailty with $\mathbb E (Z) = 1$ and $\mathbb{VAR}(Z)=\sigma^2$, the force of mortality of the population, i.e., the marginal hazard is

\begin{equation} \label{eq:ggomp}
    \bar{\mu}(x) = \frac{ae^{bx}}{1+\sigma^2 \frac{a}{b}\left( e^{bx}-1\right)}
\end{equation}
(see \cite{vaupel1979impact} and \cite{VauMis14} for all technicalities). Note that the variance of $Z$ is often denoted by $\gamma$ \citep[e.g., in ][]{VauMis14} because it is also equal to the squared coefficient of variation of the distribution of frailty among survivors to any age $x$. If $\sigma^2 > 0$, the force of mortality for the population $\bar{\mu}(x)$ starts deviating from the exponential pattern with increasing $x$ and reaches an asymptote $b/\sigma^2$. When $\sigma^2 = 0$, i.e., when there is no unobserved heterogeneity, the model for the population reduces to the (Gompertz) model for individuals with an exponentially increasing hazard function $\mu(x) = ae^{bx}$. 

Testing for mortality deceleration in this setting reduces to statistical testing whether $\sigma^2 = 0$ given the alternative $\sigma^2 > 0$. The frailty parameter $\sigma^2$ can take a value on the boundary of the parameter space ($\sigma^2 = 0$). This violates the standard underlying assumptions about the asymptotic properties of likelihood-based inference and statistical hypothesis testing \citep[see, for example,][]{bohnstedt2019detecting}. As a result, the asymptotic distribution of the maximum likelihood estimator may not be Gaussian.

In this paper, we treat the problem of identifying whether $\sigma^2 > 0$ or $\sigma^2 = 0$ as a model misspecification problem, i.e., we consider the gamma-Gompertz model when it is the Gompertz model that actually holds. In this setting, we suggest subtracting a penalty from the log-likelihood function. This penalty will be responsible for shrinking $\sigma^2$ to zero when there is no heterogeneity, as well as for adding a small bias to the Maximum Likelihood Estimator (MLE) when the effect of unobserved heterogeneity is non-negligible. We carry out Monte Carlo simulation experiments to evaluate the accuracy and precision of the estimates obtained by maximizing the likelihood function, on the one hand, and the penalized likelihood function, on the other.

In Section 2, we formulate the model misspecification problem and introduce inference methodology taking advantage of the maximum a posteriori probability (MAP). Then we carry out a Monte Carlo simulation study to compare the performance of maximizing a standard and a penalized likelihood. In Section 3, we compare the latter on mortality data for France, Japan and the USA. Section 4 discusses the advantages and drawbacks of applying our method to detect heterogeneity (deceleration) in mortality patterns.


\section{Methodology}
%
Suppose $\bm X$ is a random sample with a cumulative distribution function $G(x)$, and we fit the incorrect family of densities $\{ f(x; \bm \theta),  \bm \theta \in \Theta\}$ to the data using MLE. The misspecified log-likelihood is

\begin{equation*}
    \ell (\bm \theta; \bm X) = \sum_{i=1}^n \log f(X_i; \bm \theta).
\end{equation*}
Applying the law of large numbers, we get in the limit what the misspecified log-likelihood function $\ell (\bm \theta; \bm X)$ looks like for each $\bm \theta \in \Theta$ (see the right-hand side below):

\begin{equation} \label{eq:mle_convergence}
    \frac{1}{n}\ell (\bm \theta; \bm X) = \frac{1}{n}\sum_{i=1}^n \log f(X_i; \bm \theta) \xrightarrow[]{a.s} \mathbb E _g \left( \log f(X_1; \bm \theta)\right) = \int_{\text{Im}(X_1)} \log f(x; \bm \theta) dG(x) \,.
\end{equation}


Assume there is no heterogeneity in the data ($\sigma^2 = 0$), and we fit a gamma-Gompertz model. In other words, we observe an exponential death-rate increase in the data, but we estimate a model that implies a downward deviation from the exponential at the oldest ages. As shown in (\ref{eq:mle_convergence}), we will estimate $\sigma^2$ close to but never equal to zero. 


In this model setting, the standard technique is to estimate both the Gompertz and the gamma-Gompertz models and compare their goodness of fit. However, minor changes in the data can result in different models being selected, which can reduce prediction accuracy and lead to misinterpretations about the mortality deceleration and the mortality plateau. \cite{bohnstedt2019detecting} derive the asymptotic distribution of the likelihood ratio test statistic to detect heterogeneity. Here, we would like to suggest an alternative that does not involve hypothesis testing. Using the latter has been widely discussed and rethought in the Statistics community \citep{berk2010statistical, head2015extent, vidgen2016p, bruns2016p}, especially in relation to the arbitrary choice of the $\alpha$-level (most often 0.1, 0.05, or 0.01) and sample size issues.

Maximum likelihood estimators, obtained by maximizing the log-likelihood function, often have low bias and large variance. Estimation accuracy can sometimes be improved by shrinking some parameters to zero \citep{lasso_ridge}. The associated shrinkage estimator improves the overall prediction accuracy at the expense of introducing a small bias to reduce the variance of the parameters. This class of estimators is implicit in Bayesian inference and penalized likelihood inference. Using shrinkage estimators is applied as an alternative to hypothesis testing. Lasso, Ridge and Stein-type estimators are the most widely used examples of penalizing methods \citep[see, for example, ][]{hastie2009elements}.




\subsection{Inference}
Let $D_x$ be the number of deaths in a given age interval $[x,x+1)$ for $x=0,\ldots,m$, and $E_x$ denote the number of person-years lived in the same interval \citep[see, for example][]{brillinger1986biometrics, macdonald2018modelling}. Define $\bm D =(D_0, D_1, \dots, D_m)^\top$ and $\bm E =(E_0, E_1, \dots, E_m)^\top$. In addition, let $\bm{\theta}=(a, b, \sigma^2)^\top \in \Theta$ be the parameter vector that characterizes the force of mortality at age $x$ of the gamma-Gompertz model given by (\ref{eq:ggomp}).

Assume $D_x$ are Poisson-distributed with $ \mathbb E (D_x) = \mathbb{VAR}(D_x) = \mu (x; \bm \theta) E_x$ for $x=0,\ldots,m$ \citep{brillinger1986biometrics}. Under this assumption, the log-likelihood function for $\bm{\theta}=(a, b, \sigma^2)^\top$ is given by

\begin{equation} \label{eq:log_likelihood}
    \ell(\bm \theta ) = \ell(\bm \theta |\bm D,\bm E ) = \sum_{x=0}^m \left[D_x\ln\mu(x; \bm{\theta}))-E_x \, \mu(x; \bm{\theta})\right].
\end{equation}
Maximizing $\ell(\bm \theta )$ with respect to $\bm{\theta}=(a,b,\sigma^2)^\top$ yields the maximum-likelihood (ML) estimate $\hat{\bm \theta}$.

Let us now define a penalized log-likelihood function as

\begin{equation} \label{eq:penalized_likelihood}
    \ell_p(\bm \theta ) = \ell(\bm \theta ) - p(\sigma^2) \,, 
\end{equation}
where $\ell(\bm \theta)$ is the standard log-likelihood (\ref{eq:log_likelihood}), while $p(\sigma^2)$ is a penalty function. The penalized maximum-likelihood estimate is obtained by maximizing $\ell_p(\bm \theta )$ with respect to $\bm{\theta}=(a,b,\sigma^2)^\top$. For the problem addressed in this paper, the penalty function $p(\sigma^2)$ must be a non-decreasing monotonic continuous function and $\displaystyle{\lim_{\sigma^2 \downarrow 0} p(\sigma^2)} > p(\sigma^2)$ for all $\sigma^2 > 0$. 

In a Bayesian framework, maximizing (\ref{eq:penalized_likelihood}) is equivalent to maximizing a posterior distribution in a setting, in which $\e^{- p(\sigma^2) }/C_p$, $ C_p := \int_{\Theta} \e^{- p(\sigma^2) } \nabla \bm \theta < \infty$, is taken as a prior distribution of $\bm \theta$. This procedure yields the maximum a posteriori probability (MAP) estimator. MAP is the only Bayesian estimator that minimizes the expected canonical loss \citep{pereyra2019revisiting} and is widely used in image and video processing \citep{greig1989exact, afonso2010augmented, belekos2010maximum}.

As $\sigma^2$ describes the variance of frailty at the starting age of analysis, the standard approach would be to specify an inverse gamma prior distribution for it \citep{gelman1995bayesian}. The inverse gamma distribution is heavy-tailed and keeps probability mass further from zero than the gamma distribution. In addition, while the inverse-gamma mode is always positive, the gamma mode can also be zero \citep{llera2016estimating}. As we aim to test whether $\sigma^2 = 0$ or $\sigma^2 > 0$, we will use the log-kernel of the gamma distribution to define the penalty function as
\begin{equation}
    p(\sigma^2) = \lambda \left ( \sigma^2 + \ln\sigma^2 \right) \label{eq:penalty_function}
\end{equation}
for some non-negative $\lambda$. When $\lambda<1$, using (\ref{eq:penalty_function}) is equivalent to specifying a gamma prior distribution for $\sigma^2$ with parameters $\alpha = 1-\lambda$ and $\beta = \lambda$. 

When $m \to \infty$, the effect of the penalty diminishes regardless of the size of $\lambda$. For human life table data $m$ is finite, thus  $\lambda \geq 0$ is a constant that controls the relative impact of the penalty function on the estimates. When $\lambda = 0$, the penalty term has no effect, and maximizing the penalized likelihood will produce the standard maximum likelihood estimates (MLE). However, as $\lambda \rightarrow \infty$, the impact of the penalty grows, and the maximum penalized likelihood estimates for $\sigma^2$ will approach zero, providing high precision, but low accuracy. 

Choosing $\lambda$ is sensible in a wide range of applications \citep{li2009non, bhattacharya2014lasso}. Therefore, in accordance with the recommendations in \cite{li2009non}, we carry out a pilot simulation study, in which we find that choosing $\lambda = \frac{1}{2}$ provides similar precision to the one by MLE when $\sigma^2 > 0$, but better accuracy and precision when $\sigma^2 = 0$ (simulation results are presented in the next subsection). As a result, the final expression for the penalized log-likelihood we propose is
\begin{eqnarray} \label{eq:final-pen-lik}
    \ell_p(\bm \theta) = \sum_{x=0}^m \left[D_x\ln\mu(x; \bm{\theta})-E_x\,\mu(x; \bm{\theta})\right] - \frac{1}{2} \left ( \ln\sigma^2 + \sigma^2 \right).
\end{eqnarray}

From a Bayesian perspective, choosing $\lambda = \frac{1}{2}$ provides an informative prior distribution for $\sigma^2$. As for human populations we are likely to estimate $\sigma^2 < 1$ \citep{missov2013gamma}, the specified prior will provide for $\sigma^2$ a distribution with a mode equal to zero, a median equal to $0.4549$, and mean equal to $1$. Furthermore, the prior provides a probability mass of $0.6826$ in the interval $(0,1]$.

\begin{figure}[htb!]
    \centering
    \includegraphics[width=0.95\textwidth]{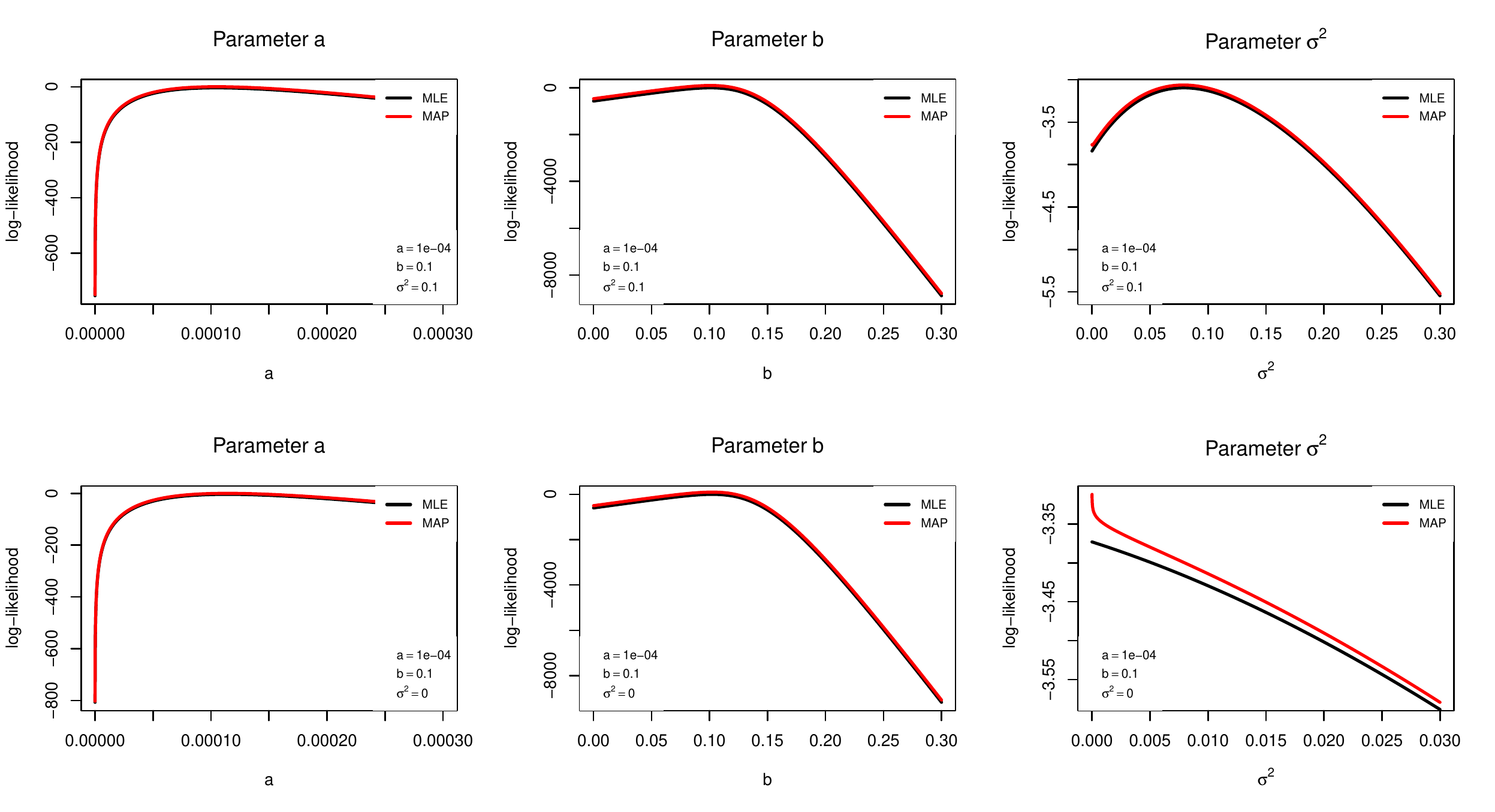}
    \caption{Plots of the profile log-likelihood (first row) and penalized log-likelihood functions (second row) of the parameters. In the first row we used synthetic data from a gamma-Gompertz model with parameters $a=0.0001$, $b=0.1$ and $\sigma^2=0.1$, in the second row we from a Gompertz model with parameters $a=0.0001$ and $b=0.1$.}
    \label{fig:likelihood_plot}
\end{figure}{}

Figure \ref{fig:likelihood_plot} shows the log-likelihood and penalized log-likelihood functions for all parameters when $\sigma^2>0$ (first row) and $\sigma^2 = 0$ (second row). When $\sigma^2>0$, the penalty function affects neither the shape of the log-likelihood nor the location of its maximum. However, when $\sigma^2 = 0$, adding a penalty yields a higher maximum at 0. Moreover, when $\sigma^2 = 0$, the first and second derivatives of the penalized log-likelihood are higher than their respective counterparts of the log-likelihood. As a result, derivative-based optimization methods may reach the maximum point faster, and the estimator $\hat{\sigma^2}$ may have a smaller variance.

\subsection{Monte Carlo simulations}
We carry out Monte Carlo simulations to explore the performance of the MAP and ML methods in estimating the gamma-Gompertz model parameters. We use the {\sf R} software \citep{R2021} to maximize the log-likelihood and the penalized log-likelihood functions via the {\tt optim} function applying as a pre-step differential evolution \citep{StoPri97, ardia2011differential}. The performance of the ML and MAP estimators are evaluated by calculating two measures: the bias and the standard deviation.

We generate 10,000 random samples from this model for some parameter values (scenarios with sample sizes of 2,000 and 5,000 were also considered, and are presented in the appendix). From these samples, we generate life tables and use them to estimate model parameters via the MAP and MLE methods. This process was repeated 2,000 times. In the presence of unobserved heterogeneity, the true parameter values are  $a_1 = 0.0001$ and $a_2 = 0.00001$ for $a$, $b_1 = 0.1$ and $b_2 = 0.15$ for $b$, and $\sigma^2_1 = 0.2$ and $\sigma^2_2 = 0.8$ for $\sigma^2$. When there is no heterogeneity ($\sigma^2 = 0$), the true parameter values are $a_1 = 0.0001$, $a_2 = 0.0003$ and $a_3 = 0.0005$ for $a$, and $b_1 = 0.09$, $b_2 = 0.10$ and $b_3 = 0.11$ for $b$. 

\begin{table}[!htp]
\centering
{\small
\caption{Simulation results: gamma-Gompertz model and sample size 10,000. }\label{tab:simulation}
\resizebox{\textwidth}{!}{%
\begin{tabular}{lcccccccccccc}   \hline
& \multicolumn{12}{c}{There is heterogeneity} \\  
\hline
& \multicolumn{6}{c}{MLE estimator} & \multicolumn{6}{c}{MAP estimator} \\ 
\hline
& \multicolumn{3}{c}{Bias} & \multicolumn{3}{c}{Standard deviation} & \multicolumn{3}{c}{Bias} & \multicolumn{3}{c}{Standard deviation} \\ 
Parameter & $a$ & $b$  & $\sigma^2$
 & $a$ & $b$  & $\sigma^2$ & $a$ & $b$  & $\sigma^2$
 & $a$ & $b$  & $\sigma^2$ \\   \hline
$(a_1, b_1, \sigma^2_1)$ & 0.000053 & -0.000051 & -0.000223 & 0.000051 & 0.001499 & 0.020721 & 0.000055 & -0.000134 & -0.001626 & 0.000052 & 0.001502 & 0.020791 \\ 
$(a_1, b_1, \sigma^2_2)$ & 0.000060 & -0.000292 & -0.007787 & 0.000056 & 0.001784 & 0.035822 & 0.000061 & -0.000354 & -0.009229 & 0.000056 & 0.001783 & 0.035795 \\ 
$(a_1, b_2, \sigma^2_1)$ & 0.000077 & 0.000131 & 0.004431 & 0.000056 & 0.002181 & 0.020569 & 0.000080 & 0.000015 & 0.003096 & 0.000057 & 0.002186 & 0.020631 \\ 
$(a_1, b_2, \sigma^2_2)$ & 0.000085 & -0.000262 & -0.003547 & 0.000061 & 0.002557 & 0.034714 & 0.000087 & -0.000348 & -0.004920 & 0.000061 & 0.002556 & 0.034687 \\ 
$(a_2, b_1, \sigma^2_1)$ & 0.000007 & -0.000349 & -0.003349 & 0.000008 & 0.001315 & 0.019464 & 0.000008 & -0.000417 & -0.004597 & 0.000008 & 0.001318 & 0.019523 \\ 
$(a_2, b_1, \sigma^2_2)$ & 0.000009 & -0.000635 & -0.013592 & 0.000008 & 0.001515 & 0.032551 & 0.000009 & -0.000683 & -0.014810 & 0.000008 & 0.001514 & 0.032528 \\ 
$(a_2, b_2, \sigma^2_1)$ & 0.000009 & -0.000170 & 0.002295 & 0.000008 & 0.001963 & 0.019377 & 0.000009 & -0.000268 & 0.001078 & 0.000008 & 0.001966 & 0.019430 \\ 
$(a_2, b_2, \sigma^2_2)$ & 0.000011 & -0.000650 & -0.007809 & 0.000009 & 0.002273 & 0.032534 & 0.000011 & -0.000721 & -0.009014 & 0.000009 & 0.002272 & 0.032510 \\ 
\hline
& \multicolumn{12}{c}{There is no heterogeneity} \\  
\hline
& \multicolumn{6}{c}{MLE estimator} & \multicolumn{6}{c}{MAP estimator} \\ 
\hline
& \multicolumn{3}{c}{Bias} & \multicolumn{3}{c}{Standard deviation} & \multicolumn{3}{c}{Bias} & \multicolumn{3}{c}{Standard deviation} \\ 
Parameter & $a$ & $b$  & $\sigma^2$
 & $a$ & $b$  & $\sigma^2$ & $a$ & $b$  & $\sigma^2(10^{-16})$
 & $a$ & $b$  & $\sigma^2(10^{-15})$ \\   \hline
$(a_1, b_1, \sigma^2)$ & 0.000005 & -0.000055 & 0.000181 & 0.000006 & 0.000875 & 0.008460 & 0.000007 & -0.000239 & 0.125942 & 0.000006 & 0.000791 & 0.6308937 \\ 
$(a_1, b_2, \sigma^2)$ & 0.000006 & -0.000050 & 0.000222 & 0.000006 & 0.000968 & 0.008907 & 0.000007 & -0.000407 & 0.060433 & 0.000006 & 0.000870 & 0.1460452 \\ 
$(a_1, b_3, \sigma^2)$ & 0.000006 & -0.000057 & 0.000428 & 0.000006 & 0.001082 & 0.009726 & 0.000008 & -0.000305 & 0.089035 & 0.000006 & 0.000978 & 8.212895 \\ 
$(a_2, b_1, \sigma^2)$ & 0.000011 & 0.000138 & 0.001864 & 0.000016 & 0.000913 & 0.009010 & 0.000016 & -0.000198 & 0.004470 & 0.000015 & 0.000811 & 0.294978 \\ 
$(a_2, b_2, \sigma^2)$ & 0.000014 & 0.000117 & 0.001094 & 0.000016 & 0.001042 & 0.009873 & 0.000019 & -0.000264 & 0.003661 & 0.000015 & 0.000859 & 4.922260 \\ 
$(a_2, b_3, \sigma^2)$ & 0.000015 & 0.000164 & 0.002204 & 0.000016 & 0.001150 & 0.010369 & 0.000022 & -0.000216 & 0.007167 & 0.000016 & 0.001017 & 3.551245 \\ 
$(a_3, b_1, \sigma^2)$ & 0.000018 & 0.000143 & 0.001062 & 0.000025 & 0.000979 & 0.009712 & 0.000027 & -0.000138 & 0.001124 & 0.000025 & 0.000876 & 1.898573 \\ 
$(a_3, b_2, \sigma^2)$ & 0.000024 & 0.000076 & 0.000505 & 0.000025 & 0.001057 & 0.009721 & 0.000030 & -0.000112 & 0.001177 & 0.000025 & 0.000942 & 8.114498 \\ 
$(a_3, b_3, \sigma^2)$ & 0.000025 & 0.000117 & 0.001427 & 0.000025 & 0.001178 & 0.009999 & 0.000033 & -0.000171 & 0.001415 & 0.000024 & 0.000990 & 0.000025 \\ 
\hline
\end{tabular}    }}
\end{table}

The simulation results are presented in Table \ref{tab:simulation}. In the presence of unobserved heterogeneity, both methods underestimate $b$ and $\sigma^2$. They also introduce a small positive bias to $a$, the one provided by ML estimator being slightly smaller. However, in general the ML and MAP estimators perform equally well, with a similar bias and standard deviation.

In the absence of unobserved heterogeneity, the ML estimator provides again a smaller bias for $a$ and $b$ than the MAP estimator. However, in this case, the MAP method estimates more precisely the frailty parameter $\sigma^2$, with a bias and a standard deviation close to zero ($\propto 10^{-15}$). The MAP estimator also provides a slight reduction in the standard deviation of parameter $b$.

By the Monte Carlo simulation we also calculate the proportion of trials in which MAP estimates $\sigma^2 > 0$ when the true values is $\sigma^2=0$ (error type I), as well as the proportion of trials in which MAP estimates $\sigma^2 = 0$ when the true values is $\sigma^2>0$ (error type II). Based on our simulations, the type I errro equals $0.001502$, while the type II error is $0.001126$.

The Monte Carlo simulations show that using a penalizing likelihood function (\ref{eq:final-pen-lik}) is an alternative to hypothesis testing, the latter being dependent on the asymptotic distribution of the ML estimator, sample size and the arbitrary choice of the $\alpha$-level \citep{bohnstedt2019detecting}.



\section{Performance of MAP and ML estimators on HMD data}
In this section, we estimate the gamma-Gompertz model via ML and MAP using mortality data from the Human Mortality Database \citep{hmd}. We take exposures and raw death counts for the female population of France, Japan and the USA in the years 1960, 1980, 2000, and 2020, after age 70. We apply again {\sf R} \citep{R2021} to compute the ML and MAP estimates of $ \bm \theta = (a, b, \sigma^2)'$ by using differential evolution. We use the mean squared error given by
\begin{equation*}
    MSE = \frac{1}{n} \sum_{x=0}^m \left(\ln m_x - \ln\bar{\mu}(x; \hat{\bm\theta})\right)^2, 
\end{equation*}
to assess the goodness of fit.

\begin{table}[!htp]
\centering
{\small
\caption{Life expectancy: gamma Gompertz–Makeham model and ML estimates.}\label{tab:estim_1}
\resizebox{\textwidth}{!}{%
\begin{tabular}{lcrrrrrrrrrrrrrr}   \hline
& & \multicolumn{4}{c}{ML Estimates} & \multicolumn{4}{c}{MAP Estimates} \\
Country & Year & $a$ & $b$  & $\sigma^2$ & MSE & $a$ & $b$  & $\sigma^2$ & MSE   \\ \hline
France & 1960 & 0.003582 & 0.107599 & 0.016726 & 0.100588 & 0.003593 & 0.107483 & 0.015902 & 0.100540\\ 
       & 1980 & 0.002210 & 0.112032 & 0.003393 & 0.045776 & 0.002220 & 0.111729 & 0.003117 & 0.046747 \\ 
       & 2000 & 0.001247 & 0.117749 & 0.000001 & 0.073648 & 0.001250 & 0.117689 & 0 & 0.073262\\ 
       & 2020 & 0.000957 & 0.119494 & 0.000002 & 0.094243 & 0.000960 & 0.119416 & 0 & 0.093697\\ \hline
Japan & 1960 &  0.004782 & 0.105858 & 0.039989 & 0.011366& 0.004782 & 0.105845 & 0.039593 & 0.011493 \\ 
       & 1980 &  0.002009 & 0.117886 & 0.015251 & 0.053944 & 0.002009 & 0.117941 & 0.015942 & 0.053328\\ 
       & 2000 &  0.001140 & 0.115268 & 0.000015 & 0.064604 & 0.001142 & 0.115118 & 0 & 0.063728\\ 
       & 2020 &  0.000575 & 0.125814 & 0.000233 & 0.104944& 0.000574 & 0.125870 & 0.000225 & 0.105597 \\  \hline
USA & 1960 &  0.004701 & 0.095797 & 0.032146 & 0.111711 & 0.004699 & 0.095814 & 0.032802 & 0.110740\\ 
       & 1980 &  0.003612 & 0.093688 & 0.000001 & 0.054483& 0.003609 & 0.093720 & 0 & 0.054652 \\ 
       & 2000 & 0.002712 & 0.100566 & 0.000003 & 0.030967 & 0.002714 & 0.100540 & 0 & 0.030873\\ 
       & 2020 & 0.002473 & 0.101652 & 0.000001 & 0.022735 & 0.002476 & 0.101612 & 0 & 0.022618\\  \hline
\end{tabular}    }}
\end{table}


Table \ref{tab:estim_1} shows the results of applying ML and MAP methods to the datasets described above. The MAP estimator provides lower MSEs in 8 of the 12 datasets. When the standard ML method estimates $\sigma^2 < 10^{-4}$, our novel method estimates $\sigma^2 = 0$ and provides a smaller MSE. This suggests that the MAP provides a slightly better fit to the data. Overall, MAP performs better than ML when unobserved heterogeneity is not detected, and while for estimates of $\hat{\sigma}^2 > 0$ ML has a slight advantage.

The results from the real-data application back up the results from the Monte Carlo simulations in Section 2. In the presence of unobserved heterogeneity, the MLE method provides the most precise and accurate estimates. The MAP method, though, has just slightly lower precision. On the other hand, in the absence of unobserved heterogeneity, the MAP provides a smaller bias and variance in its estimates compared to MLE. 

\subsection{Examples when MAP and ML estimators yield different outcomes}
Using MAP and ML estimators does not always lead to the same statistical inference. One of them can detect heterogeneity in cases when the other does not. We will illustrate this on HMD data for the Japanese female population in 2009 and the French female population born in 1848, ages 70+. To assess the goodness of fit, we will use MSE again.

For Japanese females in 2009, ML yields estimates $\hat{\bm{\theta}}_{MLE} = (0.006359, 0.133805, 0.070513)'$ with standard errors $SE(a) = 0.000188$, $SE(b) = 0.002263$ and $SE(\sigma^2) = 0.021156$. The 95\% confidence interval for $\sigma^2$ is $(0.029047, 0.111978)$ indicating stasitically significant unobserved heterogeneity, i.e., the existence of mortality deceleration. On the other hand, the MAP method estimates $\hat{\bm{\theta}}_{MAP} = (0.006966, 0.125440, 0)'$, indicating the absence of unobserved heterogeneity. Comparing the goodness of fit of both methods speaks in favor of the MAP outcome: MAP's MSE is by 37\% lower than ML's LSE (0.018691 for MAP vs 0.029958 for ML). It indicates that unobserved heterogeneity is negligible and that the gamma-Gompertz model is misspecified. 

\begin{figure}[htb!]
    \centering
    \includegraphics[width=0.95\textwidth]{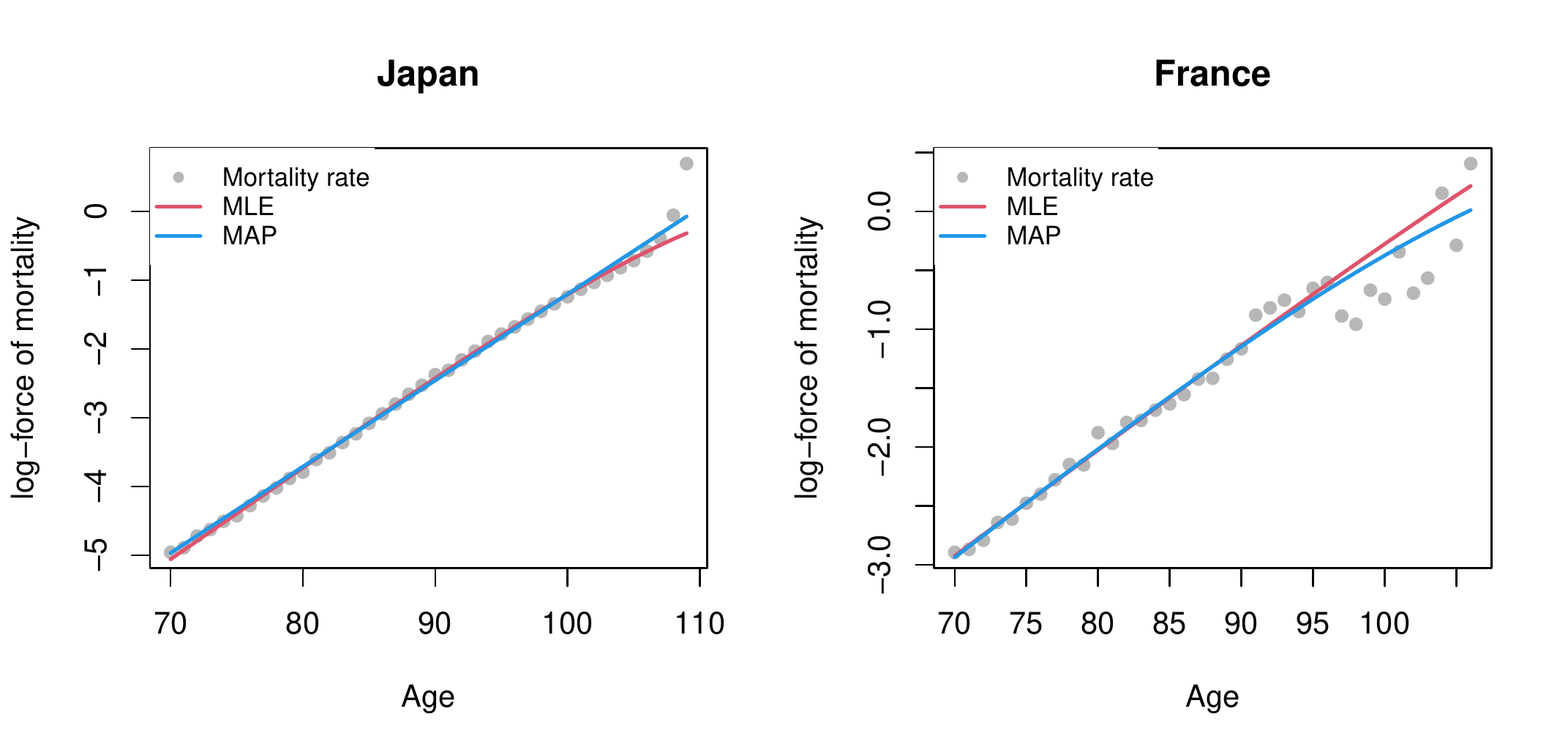}
    \caption{MAP and MLE estimates of the force of mortality for the Japanese population in 2009 and the Swedish population born in 1881, after age 70}
    \label{fig:example_fit}
\end{figure}{}

The left panel of Figure \ref{fig:example_fit} shows that both methods estimate a similar logarithmic force of mortality at most ages. However, after age 100, the MLE deviates downward from the observed logarithmic death rates. 

The MAP also provides a better fit and different conclusion for the cohort of French females born in 1848. While ML estimates $\hat{\bm{\theta}}_{MLE} = (0.053748, 0.090552, 0.008604)'$ with $SE(a) = 0.000317$, $SE(b) = 0.001273$, $SE(\sigma^2) = 0.007562$ and provides an MSE equal 0.046222, MAP estimates $\hat{\bm{\theta}}_{MAP} = (0.053113, 0.094921, 0.036466)'$ and provides $MSE = 0.034226$, i.e., MAP's MSE  is by 26\% smaller than ML's MSE. 

Furthermore, while the MAP estimate of $\sigma^2$ suggests that there is non-negligible unobserved heterogeneity, the ML estimate and standard error for $\sigma^2$ indicates the opposite: the amount of unobserved heterogeneity is not statistically significant. The right panel of Figure \ref{fig:example_fit} shows the difference between these estimates. MAP's estimate shows a leveling-off in the force of mortality, while the MLE shows a log-linear increase in the hazard function.

\subsection{Comparison between MAP and ML estimators for different populations}
To evaluate and compare empirically the performance of the ML and MAP methods we are going to use them to estimate the force of mortality for the male and female populations of France, Denmark, Sweden, Italy, Japan, Czechia, and the United States of America from 1950 to 2019, resulting in 980 populations. To access the goodness of fit we are using the MSE.

\begin{figure}[htb!]
    \centering
    \includegraphics[width=.95\textwidth]{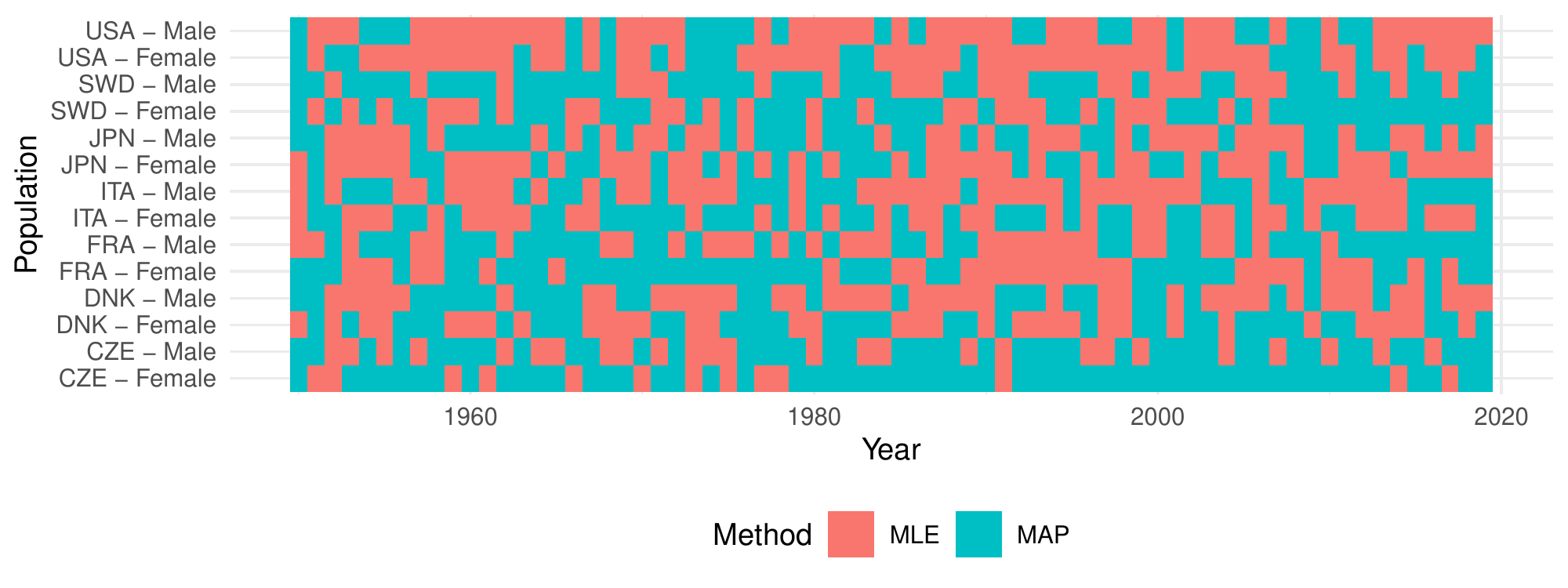}
    \caption{Method providing the best goodness of fit.}
    \label{fig:heatmap_period}
\end{figure}

Figure \ref{fig:heatmap_period} presents the method that provides a better fit (which has the smallest MSE). Overall both methods provide similar goodness of fit, with MAP providing a slightly lower MSE (on average 0.5\% smaller than the ML method). Over the 980 populations, the MAP provides a better fit for 502 of them. For Czechia, Sweden, and France, the MAP also provided a slightly better fit than the ML method; however, the MSE difference is small within the country.

{}

\begin{figure}[htb!]
    \centering
    \includegraphics[width=.95\textwidth]{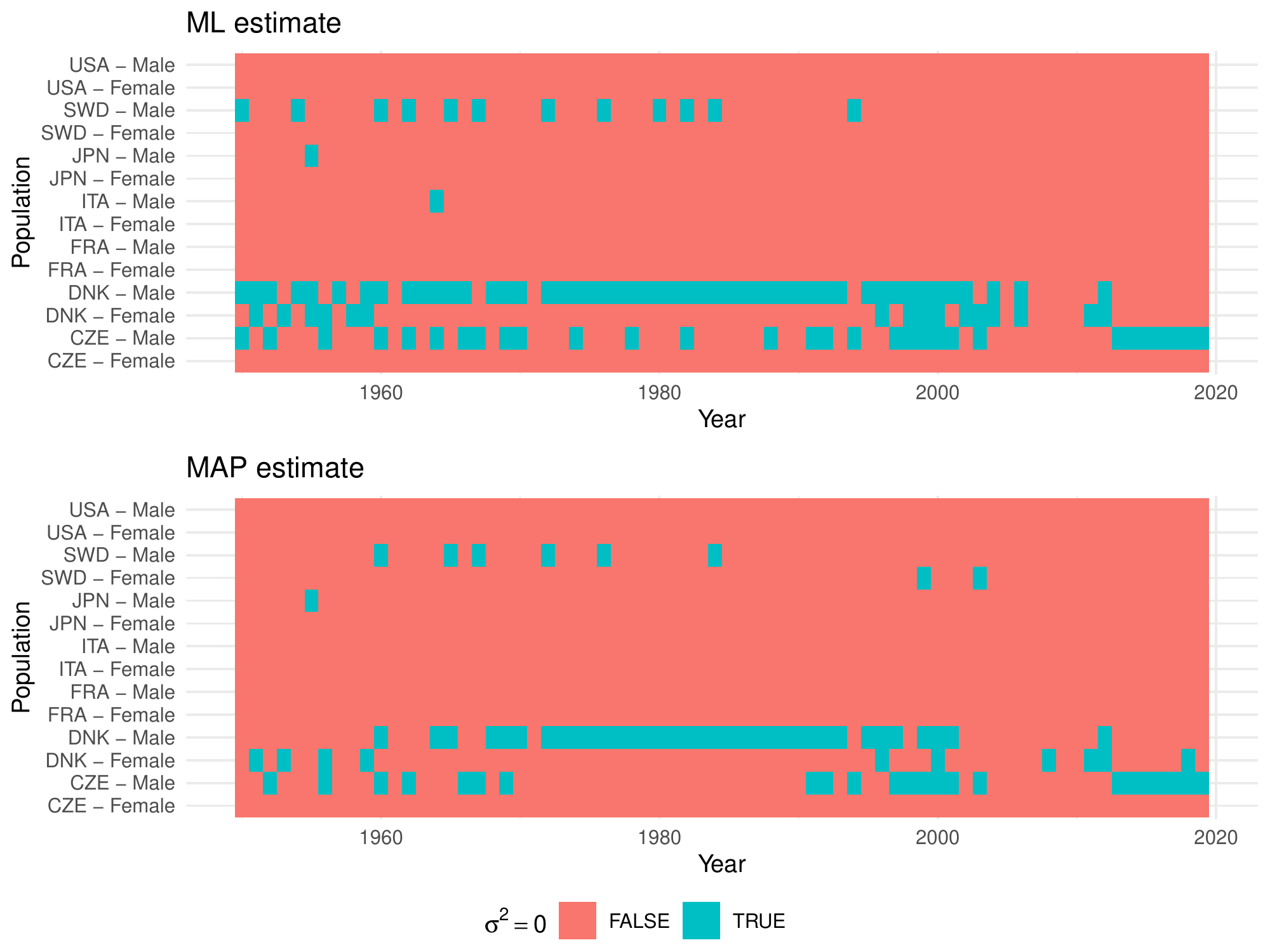}
    \caption{Statistical non-significance of $\sigma^2$}
    \label{fig:heatmap_sigma2_period}
\end{figure}

We also estimate the standard error for the ML estimates of $\sigma^2$ to test if the parameter is not statistically significantly different from zero. The results were compared with the MAP estimate. Figure \ref{fig:heatmap_sigma2_period} presents for which populations $\sigma^2$ is zero through the ML and MAP method. From both methods, it is clear that the non-estimation of unobserved heterogeneity is related to the male populations - especially for the Danes. It might be caused by the small male population surviving after age 105.

In about 96\% of the populations, the methods agree on the significance or non-significance of $\sigma^2$. However, when the MAP estimates $\sigma^2 = 0$ the ML provided positive confidence intervals for about 5.1948\% of the cases, which corresponds to the chosen $\alpha$-level (probability of Type I error) of 5\% for those intervals.

{}
\section{Concluding remarks}
 
\cite{bohnstedt2019detecting} introduced a formal procedure to identify whether $\sigma^2 > 0$ or $\sigma^2 = 0$ in a hypothesis testing setting: they studied the asymptotic properties of the maximum likelihood estimator and the likelihood ratio test (LRT) for $H_0:\sigma^2 = 0$ vs. $H_1:\sigma^2 = 0$ for the gamma-Gompertz model. However, LRTs are based on the asymptotic distribution of the maximum likelihood estimator, hence its convergence depends on the sample size. Moreover, conclusions drawn from hypothesis tests are dependent on the arbitrary choice of the significance level or $p$-value.

We suggest an alternative method by considering the problem as model misspecification. We add a penalty function to the likelihood so that we make sure that $\hat{\sigma}^2=0$ when there is no heterogeneity. We also present a Bayesian interpretation (MAP) to our method. We take advantage of robust Monte Carlo simulations to measure the bias and standard deviation of the ML and MAP methods in scenarios with and without unobserved heterogeneity. We also compare the performance of both methods for estimating the gamma-Gompertz model parameters using actual mortality data from the Human Mortality Database. The two methods work almost equally well, the ML having a slight advantage, in the presence of unobserved heterogeneity. However, in the absence of the latter, the MAP method provides an estimate closer to 0 ($\hat{\sigma}^2 \approx 10^{-20}$) and a better fit to the model in comparison to ML. As a result, the method we propose here can be used as an alternative to likelihood ratio testing for the gamma-Gompertz model with $H_0:\sigma^2 = 0$ vs. $H_1:\sigma^2 > 0$. On the one hand, the MAP method does not depend on any asymptomatic distribution, its performance is not strongly affected by sample size, and it also does not depend on the arbitrary choice of the significance level. On the other hand, MAP provides similar estimates to the ones by ML when $\sigma^2 > 0$ and more accurate estimates when $\sigma^2 = 0$.


\section*{Acknowledgments}

The research leading to this publication is a part of a project that has received funding from the European Research Council (ERC) under the European Union’s Horizon 2020 research and innovation programme (Grant agreement No. 884328 – Unequal Lifespans). Silvio C. Patricio gratefully acknowledges the support provided from AXA Research Fund, through the funding for the “AXA Chair in Longevity Research”. 

\medskip

\bibliographystyle{apalike}
\bibliography{mybibliography}

\begin{appendices}

\begin{table}[!htp]
\centering
{\small
\caption{Simulation results: gamma-Gompertz model and sample size 2,000. }\label{tab:simulation_1000}
\resizebox{\textwidth}{!}{%
\begin{tabular}{lcccccccccccc}   \hline
\multicolumn{13}{c}{There is heterogeneity} \\  
\hline
& \multicolumn{6}{c}{MLE estimator} & \multicolumn{6}{c}{MAP estimator} \\ 
\hline
& \multicolumn{3}{c}{Bias} & \multicolumn{3}{c}{Standard deviation} & \multicolumn{3}{c}{Bias} & \multicolumn{3}{c}{Standard deviation} \\ 
Parameter & $a$ & $b$  & $\sigma^2$
 & $a$ & $b$  & $\sigma^2$ & $a$ & $b$  & $\sigma^2$
 & $a$ & $b$  & $\sigma^2$ \\   \hline
$(a_1, b_1, \sigma^2_1)$ & 0.000089 & -0.001172 & -0.020361 & 0.000111 & 0.003286 & 0.047684 & 0.000104 & -0.001709 & -0.029261 & 0.000117 & 0.003465 & 0.051082 \\ 
$(a_1, b_1, \sigma^2_2)$ & 0.000110 & -0.002075 & -0.048634 & 0.000128 & 0.003972 & 0.078621 & 0.000118 & -0.002397 & -0.055984 & 0.000129 & 0.003970 & 0.078467 \\ 
$(a_1, b_2, \sigma^2_1)$ & 0.000104 & -0.000906 & -0.008883 & 0.000124 & 0.004826 & 0.047186 & 0.000120 & -0.001603 & -0.016820 & 0.000130 & 0.005032 & 0.049798 \\ 
$(a_1, b_2, \sigma^2_2)$ & 0.000124 & -0.001978 & -0.029154 & 0.000139 & 0.005738 & 0.077069 & 0.000133 & -0.002420 & -0.036104 & 0.000140 & 0.005733 & 0.076901 \\ 
$(a_2, b_1, \sigma^2_1)$ & 0.000030 & -0.003810 & -0.046694 & 0.000021 & 0.003024 & 0.045092 & 0.000034 & -0.004407 & -0.057147 & 0.000024 & 0.003388 & 0.052275 \\ 
$(a_2, b_1, \sigma^2_2)$ & 0.000038 & -0.005056 & -0.097144 & 0.000023 & 0.003338 & 0.070146 & 0.000039 & -0.005316 & -0.103429 & 0.000024 & 0.003333 & 0.069973 \\ 
$(a_2, b_2, \sigma^2_1)$ & 0.000027 & -0.003998 & -0.030152 & 0.000021 & 0.004259 & 0.043531 & 0.000030 & -0.004673 & -0.038167 & 0.000022 & 0.004481 & 0.046813 \\ 
$(a_2, b_2, \sigma^2_2)$ & 0.000032 & -0.005191 & -0.065676 & 0.000025 & 0.005158 & 0.075375 & 0.000034 & -0.005565 & -0.071836 & 0.000025 & 0.005146 & 0.075141 \\ 
\hline
\multicolumn{13}{c}{There is no heterogeneity} \\  
\hline
& \multicolumn{6}{c}{MLE estimator} & \multicolumn{6}{c}{MAP estimator} \\ 
\hline
& \multicolumn{3}{c}{Bias} & \multicolumn{3}{c}{Standard deviation} & \multicolumn{3}{c}{Bias} & \multicolumn{3}{c}{Standard deviation} \\ 
Parameter & $a$ & $b$  & $\sigma^2$
 & $a$ & $b$  & $\sigma^2$ & $a$ & $b$  & $\sigma^2(10^{-12})$
 & $a$ & $b$  & $\sigma^2(10^{-12})$ \\   \hline
$(a_1, b_1, \sigma^2)$ & 0.000015 & -0.001225 & 0.000002 & 0.000013 & 0.001615 & 0.006546 & 0.000016 & -0.001341 & 0.021800 & 0.000014 & 0.001693 & 0.000287 \\ 
$(a_1, b_2, \sigma^2)$ & 0.000014 & -0.001224 & 0.000002 & 0.000013 & 0.001833 & 0.007389 & 0.000016 & -0.001312 & 0.018541 & 0.000014 & 0.001858 & 0.004010 \\ 
$(a_1, b_3, \sigma^2)$ & 0.000014 & -0.001323 & 0.000002 & 0.000013 & 0.001998 & 0.008857 & 0.000015 & -0.001324 & 0.024195 & 0.000015 & 0.002098 & 0.001130 \\ 
$(a_2, b_1, \sigma^2)$ & 0.000028 & -0.000591 & 0.000003 & 0.000031 & 0.001703 & 0.009242 & 0.000031 & -0.000983 & 0.000923 & 0.000033 & 0.001666 & 0.000003 \\ 
$(a_2, b_2, \sigma^2)$ & 0.000029 & -0.000837 & 0.000003 & 0.000031 & 0.001857 & 0.010440 & 0.000031 & -0.000842 & 0.000731 & 0.000031 & 0.001804 & 0.000009 \\ 
$(a_2, b_3, \sigma^2)$ & 0.000030 & -0.000810 & 0.000004 & 0.000031 & 0.002067 & 0.011604 & 0.000035 & -0.001196 & 0.000701 & 0.000034 & 0.002086 & 0.000008 \\ 
$(a_3, b_1, \sigma^2)$ & 0.000034 & -0.000398 & 0.000003 & 0.000046 & 0.001739 & 0.011618 & 0.000037 & -0.000606 & 0.000220 & 0.000048 & 0.001681 & 0.000001 \\ 
$(a_3, b_2, \sigma^2)$ & 0.000036 & -0.000503 & 0.000005 & 0.000049 & 0.002062 & 0.013431 & 0.000040 & -0.000670 & 0.000151 & 0.000048 & 0.001901 & 0.000002 \\ 
$(a_3, b_3, \sigma^2)$ & 0.000040 & -0.000238 & 0.000008 & 0.000050 & 0.002247 & 0.014168 & 0.000045 & -0.000722 & 0.000108 & 0.000051 & 0.002061 & 0.000008 \\ 
\hline
\end{tabular}    }}
\end{table}
\begin{table}[!htp]
\centering
{\small
\caption{Simulation results: gamma-Gompertz model and sample size 5,000. }\label{tab:simulation_1000}
\resizebox{\textwidth}{!}{%
\begin{tabular}{lcccccccccccc}   \hline
\multicolumn{13}{c}{There is heterogeneity} \\  
\hline
& \multicolumn{6}{c}{MLE estimator} & \multicolumn{6}{c}{MAP estimator} \\ 
\hline
& \multicolumn{3}{c}{Bias} & \multicolumn{3}{c}{Standard deviation} & \multicolumn{3}{c}{Bias} & \multicolumn{3}{c}{Standard deviation} \\ 
Parameter & $a$ & $b$  & $\sigma^2$
 & $a$ & $b$  & $\sigma^2$ & $a$ & $b$  & $\sigma^2$
 & $a$ & $b$  & $\sigma^2$ \\   \hline
$(a_1, b_1, \sigma^2_1)$ & 0.000067 & -0.000379 & -0.004470 & 0.000071 & 0.002078 & 0.029046 & 0.000071 & -0.000552 & -0.007370 & 0.000072 & 0.002090 & 0.029281 \\ 
$(a_1, b_1, \sigma^2_2)$ & 0.000070 & -0.000582 & -0.015178 & 0.000077 & 0.002506 & 0.049888 & 0.000074 & -0.000708 & -0.018078 & 0.000077 & 0.002504 & 0.049808 \\ 
$(a_1, b_2, \sigma^2_1)$ & 0.000090 & -0.000273 & 0.001254 & 0.000077 & 0.002969 & 0.028061 & 0.000095 & -0.000509 & -0.001483 & 0.000078 & 0.002982 & 0.028253 \\ 
$(a_1, b_2, \sigma^2_2)$ & 0.000094 & -0.000514 & -0.007569 & 0.000083 & 0.003583 & 0.048230 & 0.000097 & -0.000687 & -0.010324 & 0.000084 & 0.003579 & 0.048152 \\ 
$(a_2, b_1, \sigma^2_1)$ & 0.000014 & -0.001320 & -0.014568 & 0.000011 & 0.001786 & 0.026808 & 0.000014 & -0.001465 & -0.017208 & 0.000011 & 0.001795 & 0.027016 \\ 
$(a_2, b_1, \sigma^2_2)$ & 0.000015 & -0.001650 & -0.033574 & 0.000013 & 0.002344 & 0.049087 & 0.000016 & -0.001748 & -0.036030 & 0.000013 & 0.002342 & 0.049016 \\
$(a_2, b_2, \sigma^2_1)$ & 0.000014 & -0.001299 & -0.006177 & 0.000012 & 0.002706 & 0.026681 & 0.000015 & -0.001505 & -0.008713 & 0.000012 & 0.002717 & 0.026863 \\ 
$(a_2, b_2, \sigma^2_2)$ & 0.000015 & -0.001571 & -0.019479 & 0.000014 & 0.003352 & 0.046735 & 0.000016 & -0.001714 & -0.021904 & 0.000014 & 0.003349 & 0.046666 \\ 
\hline
\multicolumn{13}{c}{There is no heterogeneity} \\  
\hline
& \multicolumn{6}{c}{MLE estimator} & \multicolumn{6}{c}{MAP estimator} \\ 
\hline
& \multicolumn{3}{c}{Bias} & \multicolumn{3}{c}{Standard deviation} & \multicolumn{3}{c}{Bias} & \multicolumn{3}{c}{Standard deviation} \\ 
Parameter & $a$ & $b$  & $\sigma^2$
 & $a$ & $b$  & $\sigma^2$ & $a$ & $b$  & $\sigma^2(10^{-12})$
 & $a$ & $b$  & $\sigma^2(10^{-12})$ \\   \hline
$(a_1, b_1, \sigma^2)$ & 0.000008 & -0.000312 & 0.000009 & 0.000008 & 0.001139 & 0.006564 & 0.000008 & -0.000441 & 0.024410 & 0.000009 & 0.001142 & 0.000058 \\ 
$(a_1, b_2, \sigma^2)$ & 0.000008 & -0.000339 & 0.000011 & 0.000009 & 0.001290 & 0.007315 & 0.000009 & -0.000497 & 0.024494 & 0.000009 & 0.001259 & 0.000179 \\ 
$(a_1, b_3, \sigma^2)$ & 0.000008 & -0.000237 & 0.000011 & 0.000009 & 0.001398 & 0.007896 & 0.000009 & -0.000483 & 0.037635 & 0.000009 & 0.001417 & 0.000985 \\ 
$(a_2, b_1, \sigma^2)$ & 0.000015 & 0.000055 & 0.000823 & 0.000021 & 0.001212 & 0.008106 & 0.000016 & -0.000102 & 0.001137 & 0.000022 & 0.001184 & 0.000012 \\ 
$(a_2, b_2, \sigma^2)$ & 0.000017 & 0.000028 & 0.000732 & 0.000022 & 0.001363 & 0.009211 & 0.000020 & -0.000228 & 0.001954 & 0.000023 & 0.001319 & 0.001225 \\ 
$(a_2, b_3, \sigma^2)$ & 0.000019 & 0.000122 & 0.000182 & 0.000022 & 0.001473 & 0.009371 & 0.000021 & -0.000024 & 0.002634 & 0.000022 & 0.001387 & 0.000014 \\ 
$(a_3, b_1, \sigma^2)$ & 0.000023 & 0.000212 & 0.001357 & 0.000033 & 0.001265 & 0.009088 & 0.000023 & -0.000068 & 0.000266 & 0.000033 & 0.001200 & 0.000001 \\ 
$(a_3, b_2, \sigma^2)$ & 0.000025 & 0.000198 & 0.000845 & 0.000034 & 0.001391 & 0.009982 & 0.000030 & -0.000154 & 0.000558 & 0.000035 & 0.001332 & 0.000048 \\ 
$(a_3, b_3, \sigma^2)$ & 0.000026 & 0.000127 & 0.000722 & 0.000035 & 0.001590 & 0.011029 & 0.000034 & -0.000151 & 0.001303 & 0.000035 & 0.001464 & 0.003727 \\ 
\hline
\end{tabular}    }}
\end{table}

\end{appendices}

\end{document}